\documentclass[12pt]{article}
\usepackage{setspace} 
\usepackage{graphicx}

\begin{document}

{\Large\bf Optical and infrared flares from a transient Galactic 
soft $\gamma$-ray repeater}

\vspace{1cm}

A. J. Castro-Tirado$^{1}$, A. de Ugarte Postigo$^{2,1}$, J. Gorosabel$^{1}$,
M. Jel\'{\i}nek$^{1}$, T. A. Fatkhullin$^{3}$, V. V. Sokolov$^{3}$,  
P. Ferrero$^{4}$, D. A. Kann$^{4}$,  S. Klose$^{4}$, D. Sluse$^{5}$, 
M. Bremer$^{6}$, J. M. Winters$^{6}$, D. Nuernberger$^{2}$,  
D. P\'erez-Ram\'{\i}rez$^{7}$,  M. A. Guerrero$^{1}$, J. French$^{8}$,
G. Melady$^{8}$, L. Hanlon$^{8}$, 
B. McBreen$^{8}$, F. J. Aceituno$^{1}$, R. Cunniffe$^{1}$, 
P. Kub\'anek$^{1}$, S. Vitek$^{1}$, S. Schulze$^{4}$, 
A. C. Wilson$^{9}$, R. Hudec$^{10}$, J. M. Gonz\'alez-P\'erez$^{11}$, 
T. Shahbaz$^{11}$, S. Guziy$^{12}$, S. B. Pandey$^{13}$
L. Pavlenko$^{14}$, E. Sonbas$^{3,15}$, S. A. Trushkin$^{3}$, 
N. N. Bursov$^{3}$, N. A. Nizhelskij$^{3}$ and L. Sabau-Graziati$^{16}$

\vspace{1cm}

$^{1}$ Instituto de Astrof\'{\i}sica de Andaluc\'{\i}a (IAA-CSIC), P.O. Box 03004, 18080 Granada, Spain.\\
$^{2}$ European Southern Observatory, Casilla 19001, Santiago 19, Chile.\\
$^{3}$ Special Astrophysical Observatory of Russian Academy of Science (SAO-RAS), Nizhnij Arkhyz, Karachai-Cherkessia, 369167 Russia.\\ 
$^{4}$ Th\"uringer Landessternwarte Tautenburg, Sternwarte 5, D-07778 Tautenburg, Germany.\\
$^{5}$ Laboratoire d\' \rm Astrophysique, \'Ecole Polytechnique F\'ed\'erale 
de Lausanne (EPFL) Observatoire, 1290 Sauverny, Switzerland.\\
$^{6}$ Institute de Radioastronomie Millim\'etrique (IRAM), 300 rue de la Piscine, 38406 Saint Martin d\' \rm H\'eres, France.\\ 
$^{7}$ Facultad de Ciencias Experimentales, Universidad de Ja\'en, Campus Las Lagunillas, E-23071 Ja\'en, Spain.\\
$^{8}$School of Physics, University College Dublin, Dublin 4, Ireland.\\
$^{9}$Department of Astronomy, University of Texas, Austin, TX 78712, USA.\\
$^{10}$Astronomical Institute of the Czech Academy of Sciences, Fricova 298,
25165 Ond\v {r}ejov, Czech Republic.\\ 
$^{11}$Instituto de Astrof\'{\i}sica de Canarias (IAC), Via L\'actea s/n, E-38205 La Laguna (Tenerife), Spain.\\
$^{12}$ Nikolaev State University, Nikolskaya 24, 54030 Nikolaev, Ukraine.\\
$^{13}$ Aryabhatta Research Institute of Observational-Sciences (ARIES), Manora Peak, NainiTal, Uttarakhand, 263129, India.\\
$^{14}$ Crimean Astrophysical Observatory, 98409 Nauchny, Ukraine.\\
$^{15}$ University of Cukurova, Department of Physics, 01330 Adana, Turkey.\\ 
$^{16}$ Instituto Nacional de T\'ecnica Aerospacial (INTA), Ctra. de Ajalvir km. 4, 28750 Torrej\'on de Ardoz (Madrid), Spain.\\

\vspace{1cm}

{\bf 
Soft  $\gamma$-ray repeaters (SGRs) are a rare type of $\gamma$-ray transient
sources that are ocasionally detected as bursts in the high-energy sky$^{1}$. 
They are thought to be produced by {\it magnetars}, young neutron
stars with very strong magnetic fields of the order of 10$^{14-15}$
G (refs 2, 3). Only three such objects are known in our Galaxy, 
and a fourth one is associated with the supernova remnant N49 in the 
Large Magellanic Cloud$^{4}$. 
In none of these cases has an optical counterpart to either the $\gamma$-ray 
flares or the quiescent source been identified. Here we present 
multi-wavelength observations of a\, puzzling source, SWIFT J195509+261406, for 
which we detected more than 40 flaring episodes in the optical band over a 
time span of 3 days, plus a faint infrared flare 11 days later, after which it 
returned to quiescence. We propose that SWIFT J195509+261406 is a member of 
a subgroup of SGRs for which the long-term X-ray emission is transient in 
nature. Furthermore, it is the first SGR for which bursts have been 
detected in the optical and near-infrared bands and maybe the link between 
the ``persistent'' SGRs and the dim isolated neutron stars.
}

On 10 June 2007, 20:52:26 UT {\it  Swift}/BAT  detected  GRB  070610  /  SWIFT
J195509+261406,  as a gamma-ray burst (GRB) with a single  ``FRED''-like  
(fast rise phase and exponential decay) profile, lasting about 8 s 
in total$^{5}$. 
Following typical procedures for GRB follow-ups, {\it Swift}/XRT began to 
observe the field shortly after the discovery and detected an X-ray 
counterpart$^{6}$. Ground based observations also detected an optical
source consistent with the high-energy detections. However, it was 
soon realized that this counterpart behaved differently from the ``classical''
 GRB afterglows  observed so  far, i.e. the light curves showed evidence  
of rapid  variability with intense flaring activity$^{7}$. The X-ray spectrum, 
obtained during the first 8000 s of observation, was modeled with an
absorbed power law, 
with a total column density consistent with the Galactic 
value of $N(H)$ = 1 $\times$ 10$^{22}$  cm$^{-2}$. 
This, together with the location of the source in the Galactic plane 
($l^{II}$ = $-$1$^{\circ}$), supports the view that the source is hosted 
by the Milky Way$^{8}$ as we show in this work. 

We triggered a  multi-wavelength observing campaign including 
radio observations at RATAN-600, millimeter observations at  Plateau de Bure, 
optical photometry at several ground-based observatories, infrared adaptive 
optics imaging at the ESO Very Large 8.2m Telescope, optical spectroscopy at 
the Russian 6.0m telescope and late time X-ray observations with 
{\it XMM-Newton}. We also carried out CO (J = 1-0) millimeter observations 
at Pico Veleta in order to search for molecular clouds towards the line of 
sight. This campaign was supplemented by available {\it Swift}/XRT data. 
For further details, see the Supplementary Material provided with this work.

Our data were collected starting $\sim$1 min after the burst trigger time. 
In the first three nights of our observations, the source displayed strong 
flaring activity$^{9}$ (Fig. 1). The flares of  SWIFT J195509+261406 (see
Table 2 of the Supplementary Material) had durations in the range of tens of
seconds to a few minutes and flux amplitudes of up to $\sim$ 10$^{2}$ with
respect to the ``outburst'' basal flux (or $\geq$ 10$^{4}$ with respect
to the quiescent state; Fig. 2).  Strong optical and X-ray flaring was 
detected over a 3 day activity period. After 13 June, the activity decayed 
abruptly (Fig. 3) and no further flares were seen until 21 June, when a 
late-time, lower-brightness flare was detected in the $H$-band using the 
8.2m VLT (+NACO).
A late-time observation by {\it XMM-Newton} $\sim$ 170 days after the burst, 
failed to 
detect the source, imposing an upper limit (3$\sigma$) to any underlying 
X-ray flux of $\leq$ 3.1  $\times$  10$^{-14}$  erg cm$^{-2}$  s$^{-1}$ 
(0.2-10 keV).

It has been suggested that the source is similar to the black 
hole candidate V4641 Sgr$^{10}$, a Galactic microquasar for which 
rapid optical variations of a factor 
up to 500 on a timescale of tens of seconds were also 
detected$^{11}$ and X-ray flares with peak luminosities of 
up to $\sim$ 4 $\times$ 10$^{39}$ erg s$^{-1}$ (ref. 12), 
far above the Eddington luminosity of a 10 M$_{\rm \odot}$ black hole. 
In fact, it has been proposed that V4641 Sgr (a high-mass 
X-ray binary) and SWIFT J195509+261406 belong to the same class of 
astrophysical objects$^{13}$.
However, several lines of evidence point against this association.
First, the lack of further detections at $\gamma$-ray (by {\it Swift}/BAT),
millimeter ($<$ 0.6 mJy, 3$\sigma$) and centimeter wavelengths 
($<$ 0.3 mJy, 3$\sigma$, ref. 14) implies a different behaviour from Galactic 
microquasars, which produce considerable gamma-ray and radio emission at 
the time of the outbursts$^{15,16}$. 
Moreover, the lack of  H$\alpha$ emission ($<$ 9.0 $\times$ 
10$^{-16}$ erg s$^{-1}$ cm$^{-2}$  within a 2$^{\prime\prime}$ aperture) 
in our spectra and in our narrow-band images obtained at the 6.0m BTA makes 
it unlikely that it is an accreting black hole candidate in a binary 
system.

Another possibility is that this source is mimicking the 
``bursting pulsar'' GRO J1744--28, a low-mass X-ray binary which 
displayed $\sim$20 hard  X-ray bursts per hour following its discovery, 
before it entered a regime of hourly bursting lasting for nearly 4 months,
with the burst rate decreasing dramatically after that time$^{17,18}$.
Flare peak luminosities reached up to $\sim$ 10$^{39}$ erg s$^{-1}$ (ref. 19), 
far above the Eddington luminosity of a 1.4 M$_{\rm \odot}$ neutron star. 
No burst episodes were reported at other wavelengths 
for this pulsar, possibly due to the high Galactic extinction 
along the line of sight. 

Contrary to GRO J1744--28, however, the fact that {\it Swift}/BAT has not 
recorded any other gamma-ray burst from SWIFT J195509+261406 after the 
initial one, points to the interesting possibility that 
SWIFT J195509+261406 is a new soft gamma-ray repeater (SGR) 
in our Galaxy, from which only the 
initial hard spike of a bursting activity period has been recorded in 
$\gamma$ rays. If this is the case, SWIFT J195509+261406 becomes the first SGR 
detected at optical wavelengths, as a previous claim of such 
a detection for SGR 0526--66 (ref. 20) could not be firmly established.
The lognormal distribution of the optical flares strengthens this association 
(see Fig. 4).

In contrast to SGR 0526--66, SGR 1806--20 and SGR 1900+14, which all show
``persistent'' X-ray emission in the range $\sim$ 10$^{-11}$ to 10$^{-12}$ 
erg cm$^{-2}$  s$^{-1}$, SWIFT J195509+261406 strongly resembles
the ``transient'' behaviour of SGR 1627--41. The latter source experienced
an activity period for 6 weeks in 1998 (ref. 21) and its underlying X-ray flux 
was observed to decrease by a factor of $\sim$ 50 over a timespan of 
1,000 days, flattening off at 
$\sim$ 3 $\times$ 10$^{-13}$  erg cm$^{-2}$  s$^{-1}$.
Based on this, we suggest that SWIFT J195509+261406 is, together 
with SGR 1627--41, a ``transient'' soft gamma-ray repeater, 
being the first SGR (either persistent
or transient) that shows strong and protracted optical flaring activity in our
Galaxy. Assuming a distance of $\sim$ 11 kpc for SGR 1627--41, 
its quiescent luminosity in the 2-10 keV band is 
L$_{\rm X}$ $\sim$ 4 $\times$ 10$^{33}$ erg s$^{-1}$ (ref. 22). 
The X-ray luminosity of SWIFT J195509+261406 during the first
8,000 s that followed the initial $\gamma$-ray spike was 
$\sim$ 4.7 $\times$ 10$^{34}$ (D/10 kpc)$^2$ erg s$^{-1}$ and 
$\leq$ 3.6 $\times$ 10$^{32}$ (D/10 kpc)$^2$ erg s$^{-1}$ at 
the time of our late-time X-ray observation after $\sim$ 170 days. 
If the {\it SWIFT} source has the same quiescent L$_{\rm X}$ as SGR 1627-41 
(the other member of the transient SGR class), then
from the limit imposed by {\it XMM-Newton} it would lie at $\geq$ 35 kpc,
 far beyond the outer spiral arm in our Galaxy.
In fact, an upper limit to the distance of D $<$ 100 kpc can be derived 
assuming that the peak of the X-ray flare observed on 11 June does not 
exceed by a factor of $\sim$ 10 the Eddington luminosity for a 
1.4 M$_{\odot}$ neutron star$^{23}$.
On the other hand, taking into account the lower limit to its distance inferred
by us from the CO observations (D $>$ 7.0 kpc, see Supplementary Material), 
we can conclude that the distance of SWIFT J195509+261406 is 
in the range 7-20 kpc (comparable to the three known SGRs in the Galaxy), 
and that its quiescent 
L$_{\rm X}$ $\leq$ (2.5-7.2) $\times$ 10$^{32}$ erg s$^{-1}$.

A deeper X-ray observation together with a detailed study of future activity 
periods of  SWIFT J195509+261406 can shed light into this new ``transient'' 
SGR class and whether it constitutes a possible link between ``persistent'' 
SGRs (with L$_{\rm X}$ $\sim$ (2-4) $\times$ 10$^{35}$ erg s$^{-1}$) 
and dim isolated neutron stars$^{24}$ 
(with L$_{\rm X}$ $\sim$ (2-20) $\times$ 10$^{30}$ erg s$^{-1}$).

\newpage

{\bf References.}\\

1. Mazets, E. P., Golenetskii, S. V., Gurian, Iu. A. \& Ilinskii, V. N. The 
5 March 1979 event and the distinct class of short gamma bursts: Are they of 
the same origin? {\it Astrophys. and Sp. Sci.} {\bf 84}, 173-189 (1982).\\

2. Kouveliotou, C. {\it et al.} The Rarity of Soft Gamma-Ray Repeaters Deduced from Reactivation of SGR 1806-20, {\it Nature} {\bf 368}, 125-127 (1994).\\

3. Thompson, C. \& Duncan, R. C. The Soft Gamma Repeaters as Very 
Strongly Magnetized Neutron Stars. II. Quiescent Neutrino, X-Ray, and 
Alfven Wave Emission, {\it Astrophys. J.} {\bf 473}, 322-342 (1996).\\

4. Hurley, K.  The 4.5$\pm$0.5 Soft Gamma Repeaters in Review, in 
{\it Gamma-Ray Bursts, 5th Huntsville Symposium, Huntsville},  
Eds. R. M. Kippen, R. S. Mallozzi, and G. J. Fishman, AIP Conf. 
Proc. {\bf 526}, 763-770 (2000).\\

5. Tueller, J., Barbier, L., Barthelmy, S. D. {\it et al.} GRB 070610, 
Swift-BAT refined analysis, {\it GCN Circ.} 6491 (2007).\\ 

6. Pagani, C. \& Kennea, J. A. GRB 070610; Swift-XRT position, 
{\it GCN Circ.} 6490 (2007).\\

7. Stefanescu, A. {\it et al.} Very fast optical flaring from a possible 
new Galactic magnetar, {\it Nature}, submitted (2008).\\ 

8. Kann, D. A., Wilson, A. C., Schulze, S. et al. GRB 070610: TLS RRM 
sees flaring behaviour - Galactic transient?, {\it GCN Circ.} 6505 (2007).\\

9. de Ugarte Postigo, A., Castro-Tirado, A. J. \& Aceituno, F. GRB 070610: 
Optical observations from OSN, {\it GCN Circ.} 6501 (2007).\\

10. Markwardt, C., Pagani, C., Evans, P. et al. SWIFT J195509.6+261406 / 
GRB 070610: A Potential Galactic Transient, 
{\it Astronomer\' \rm s Telegram} 1102 (2007).\\

11. Lindstrom, C. {\it et al.} New clues on outburst 
mechanisms and improved spectroscopic elements of the black hole binary 
V4641 Sagittarii, {\it Monthly Not. Royal Astron. Soc.} {\bf 363}, 
882-890 (2005).\\

12. Revnivtsev, M., Gilfanov, M., Churazov, E. \& Sunyaev, R. 
Super-Eddington outburst of V4641 Sgr, {\it Astron. Astrophys.} 
{\bf 391}, 1013-1022 (2002).\\

13. Kasliwal, M. M. {\it et al.} GRB 070610: A curious galactic transient, 
{\it Astrophys. J.} submitted, preprint 
available at http://arXiv.org/astro-ph/0708.0226 (2007).\\

14. Frail, D. A. \& Chandra, P.  VLA upper limit on GRB 070610, 
{\it GCN Circ.} 6539 (2007).\\ 

15. Mirabel, I. F. \& Rodr\'{\i}guez, L. F. A Superluminal Source in the Galaxy, {\it Nature} {\bf 371}, 46-48 (1994).\\

16. Hjellming, R. M. {\it et al.} Light Curves and Radio Structure of the 
1999 September Transient Event in V4641 Sagittarii 
(=XTE J1819-254=SAX J1819.3-2525), {\it Astrophys. J.} {\bf 544}, 
977-992 (2000).\\

17. Kouveliotou, C. {\it et al.} A new type of 
transient high-energy source in the direction of the Galactic Centre, 
{\it Nature} {\bf 379}, 799-801 (1996).\\

18. Finger, M. H.  {\it et al.} Discovery of hard X-ray pulsations from the 
transient source GRO J1744-28, {\it Nature} {\bf 381}, 291-293 (1996).\\

19. Sazonov, S., Sunyaev, R. \& Lund, N. Super-Eddington X-ray 
Luminosity of the Bursting Pulsar GRO J1744-28: WATCH/GRANAT Observations,
{\it Astronomy Letters} {\bf 23}, 326-334 (2005).\\
 
20. Pedersen, H. {\it et al.} Detection of possible 
optical flashes from the gamma-ray burst source GBS0526-66, {\it Nature} 
{\bf 312}, 46-48 (1984).\\

21. Woods, P. M. {\it et al.} Discovery of a new soft gamma-ray repeater, 
SGR1627-41,  {\it Astrophys. J.} {\bf 519}, L139-142 (1999).\\

22. Mereghetti, S., Esposito, P. \& Tiengo, A. XMM-Newton observations 
of soft gamma-ray repeaters, {\it Astrophys Space Sci} {\bf 308}, 
13-23 (2007).\\

23. Begelman, M. C. Super-Eddington fluxes from thin accretion disks?, 
 {\it Astrophys. J.} {\bf 568}, L97-L100 (2002).\\

24. Treves, A., Turolla, R., Zana, S. \& Colpi, M. Isolated neutron stars: 
accretors and coolers, {\it Publ. Astron. Soc. Pacific} {\bf 112}, 
297-314 (2000). \\

25. Schlegel, E. M., Finkbeiner, D. P. \& Davis, M. Maps of Dust Infrared 
Emission for Use in Estimation of Reddening and Cosmic Microwave Background 
Radiation Foregrounds,  {\it Astrophys. J.} {\bf 500}, 525-553 (1998).\\

26. The SGR/AXP online catalog provided by the McGill Pulsar Group is 
available at http://www.physics.mcgill.ca/$\sim$pulsar/magnetar/main.html . \\

27. Israel, G. {\it et al.} Discovery and monitoring of the likely IR 
counterpart of SGR 1806-20 during the 2004 $\gamma$-ray burst-active state,
{\it Astron. Astrophys.} {\bf 438}, L1-L4 (2005). \\

28. Kosugi, G., Ogasawara, R. \& Terada, H. A variable infrared counterpart 
to the soft gamma-ray repeater SGR 1806-20, {\it Astrophys. J.} {\bf 623}, 
L125-L128 (2005). \\

29. Hurley, K., McBreen, B., Delaney, M. \& Britton, A. Lognormal Properties 
of SGR 1806-20 and Implications for Other SGR Sources,  
{\it Astrophys. Space Sci.} {\bf 231}, 81-84 (1995).\\

30. G\"o\u{g}\"u\c{s}, E. {\it et al.} Statistical Properties of SGR 1900+14 
Bursts,  {\it Astrophys. J.} {\bf 526}, L93-96 (1999).\\


{\bf Acknowledgements.} 
Based on observations carried out with the IRAM Plateau de Bure
Interferometer. IRAM is supported by INSU/CNRS (France), MPG (Germany)
and IGN (Spain). We thank S. Leon for obtaining the CO (J = 1-0) data at 
Pico Veleta. 
We also thank the ESO Director's Discretionary Time Committee
for accepting the observation a few days after the onset of the event.
We are also indebted to  N. Schartel for alloting XMM-Newton DDT time to 
get a late time X-ray observation. We thank E. Alfaro for very fruitful 
conversations. P.F., D.A.K., and S.K. acknowledge financial support 
by DFG grant Kl 766/13-2.This work was partially supported by
the Spanish projects AYA 2004-01515, AYA 2007-63677 and ESP2005-07714-C03-03.

\newpage

\begin{center}
{FIGURE CAPTIONS.}
\end{center}

\vspace{0.5cm}

{\bf Figure 1. Flaring activity of SWIFT J185509+261406 on 11/12 June 2007.} 
During a 
4 hr period, the 1.2m Mercator telescope continuously observed the object
with a sequence of 42 s exposures (plus 18 s readout) with an $I_c$-band filter.
The dark area at the bottom of the plot represents the limit to any flaring 
activity of these observations.
The brightest flare in this period (inset) reaches $I_c$ $\sim$ 17.8, 
which is still three magnitudes dimmer than the brightest flare, detected 
with WATCHER (reaching $I_c$ $\sim$ 14.7) which 
corresponds to a peak flux density of 2.96 mJy 
(not de-reddened, or 0.68 Jy de-reddened, for  A$_{\rm I}$ = 5.9, ref. 25) 
A late-time flare (not shown here) was observed with the 8.2m VLT telescope on 
22 June 2007 (08:09:27 UT). 
By analysing the different optical frames, we derive 
the following position for the source ($\pm$ 0$^{\prime\prime}$.27): 
R.A. (J2000) = 19h 55m 09.653s, 
Decl.:(J2000) = +26$^{\circ}$ 14$^{\prime}$ 05$^{\prime\prime}$.84.

\vspace{0.5cm}

{\bf Figure 2. Optical and X-ray light curves of 
SWIFT J195509+261406 (June-November 2007)}. 
{\bf a,} Optical detections ($I_c$-band magnitudes, 
filled circles, with 1$\sigma$ s.d. error bars) are shown together 
with upper limits (3$\sigma$, downward pointing triangles).  
{\bf b,} {\it Swift} X-ray data (0.2-10 keV, filled circles, 
with 1$\sigma$ s.d. error bars) together with the late time limit 
(3$\sigma$) obtained with {\it XMM-Newton}. Both light
curves show a strong activity during the first three days, reaching the 
maximum around one day after the bursts and gradually decaying
after the third day until the source became undetectable.
The X-ray observations performed by {\it Swift} do not overlap with the 
times of any of the optical flares we have recorded. 
However, observations in both X-ray and optical agree that the strongest 
flaring activity is found around one day after the $\gamma$-ray event. 
A powerful 
X-ray flare, for which the flux increased by $\sim$100 in a timescale of 
less than 10$^4$ s, was followed by several optical flares 
of similar amplitude.

\vspace{0.5cm}

{\bf Figure 3. Deep, late observations of the SWIFT J195509+261406 field.} 
{\bf a,} Deep $I_c$-band image obtained with the 6.0m BTA (+SCORPIO) 
obtained on 12 October 2007. 
{\bf b,} Deep $H$-band image obtained on 30 Sep 2007 with the 
8.2m VLT (+ NACO) using laser guide star adaptive optics. 
Both images show that the source has disappeared. 
The location for SWIFT J185509+261406, is marked with a circle (error radius
of 0$^{\prime\prime}$.27). 
The limiting magnitudes are $I_c$ $>$ 23.5 and $H$ $>$ 23.0. 
Four anomalous X-ray pulsars (AXPs), a subclass of magnetars, have been 
detected at near-IR wavelengths$^{26}$ but no $H$-band 
counterpart of any SGR is known. SGR 1806--20, which is hidden by more 
than 30 mag visual extinction, was only seen in the $K$-band when it was 
in an active state$^{27,28}$. All other three SGRs known so far have no NIR 
counterpart at all.
\vspace{0.5cm}

{\bf Figure 4. Lognormal distribution of flare fluxes for 
SWIFT J195509+261406.} 
The magnitude distribution of the optical flares 
detected from SWIFT J195509+261406 in the $I_c$ band is shown. 
Using all $I_c$-Band detections of the source,
we find that the flare fluxes are lognormally distributed as
seen in the high-energy flares of SGR 1806--20 (ref. 29) 
and SGR 1900+14 (ref. 30), supporting the belief that 
SWIFT J195509+261406 is a new SGR. 
We find that the 
observed data is well fit by a truncated normal distribution 
(solid line) with $N=\frac{A}{w\sqrt{\pi/2}}e^{-2\frac{(x-
x_c)2}{w^2}}$. Here, $N$ is the number of flares in a one 
magnitude bin, $x$ is the magnitude, $x_c$ is the center of the 
distribution, $w$ is the width and $A$ is the amplitude. The fit 
is moderately well acceptable with $\chi^{2}$=49.7 for 35 
degrees of freedom, and we find as parameters: 
$x_c=20.80\pm0.61$, $w=2.96\pm0.76$ and 
$A=64.5\pm22.7$. The truncation of the distribution is a natural 
result of the limiting magnitude of the observations. Such a 
lognormal distribution of fluxes is typical for high-energy bursts 
from SGRs.

\newpage

\begin{figure}
\includegraphics[width=13.5cm,angle=0,clip=true]{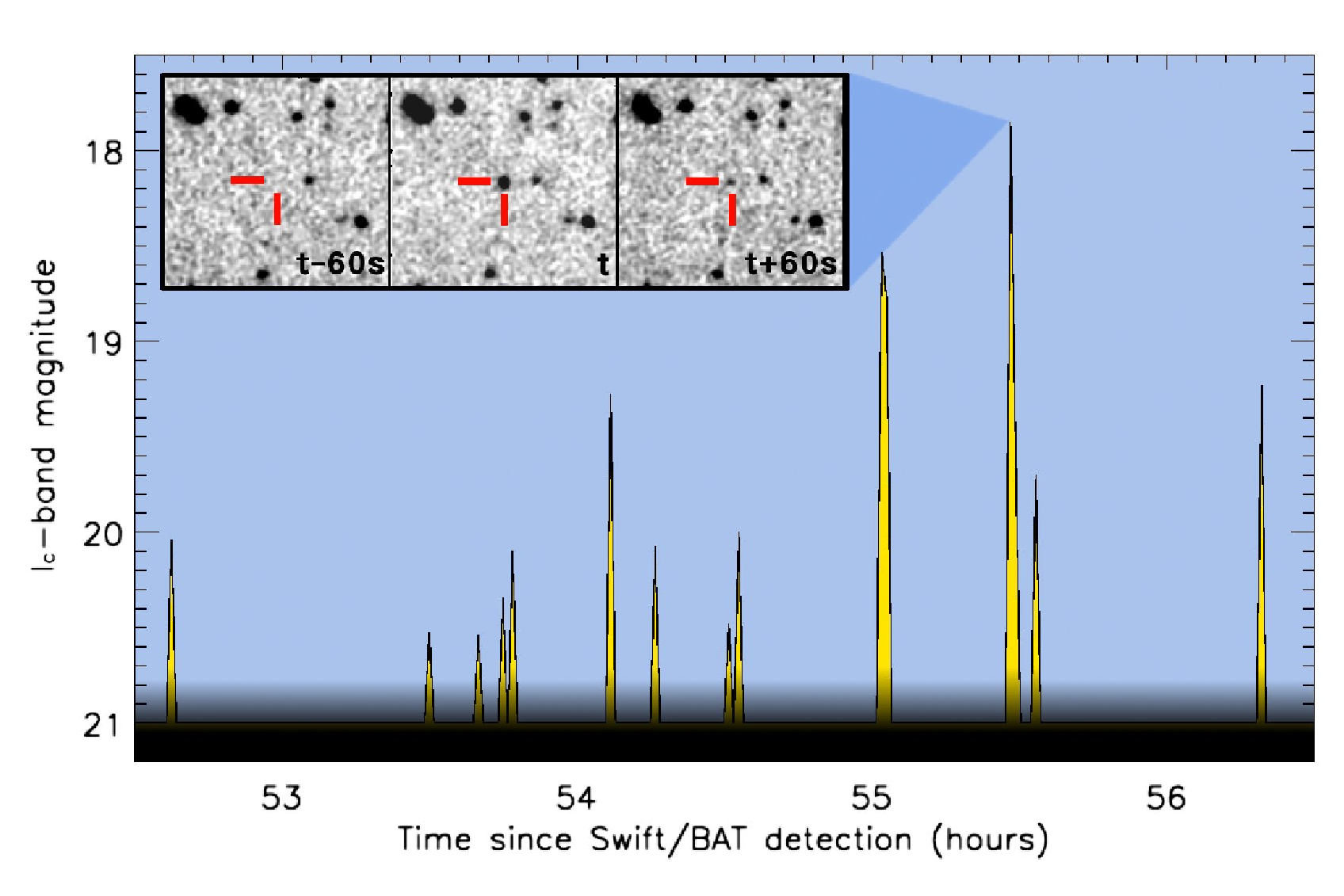}
\caption{}
\end{figure}

\begin{figure}
\includegraphics[width=13.5cm,angle=0,clip=true]{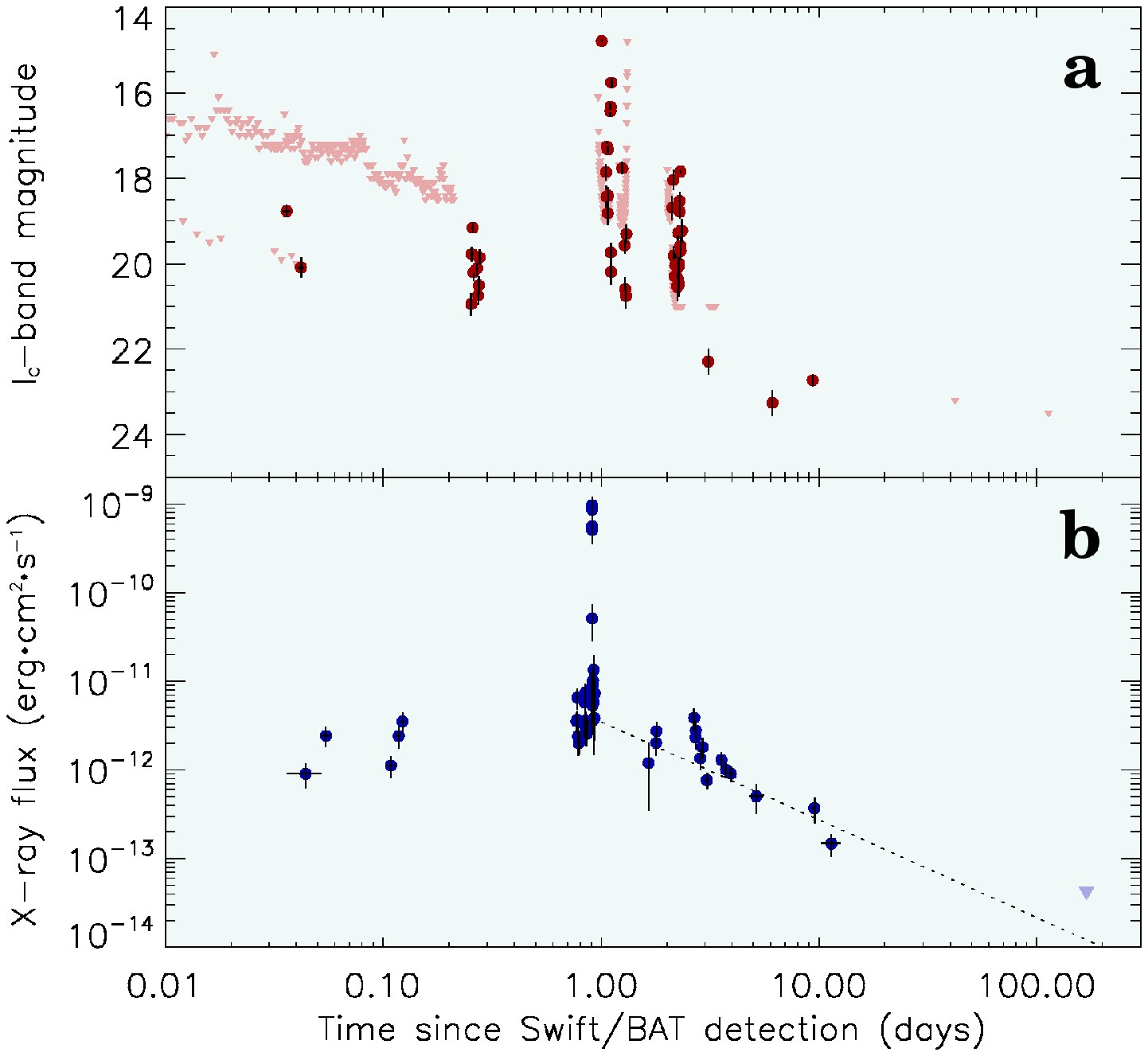}
\caption{}
\end{figure}

\begin{figure}
\includegraphics[width=13.5cm,angle=0,clip=true]{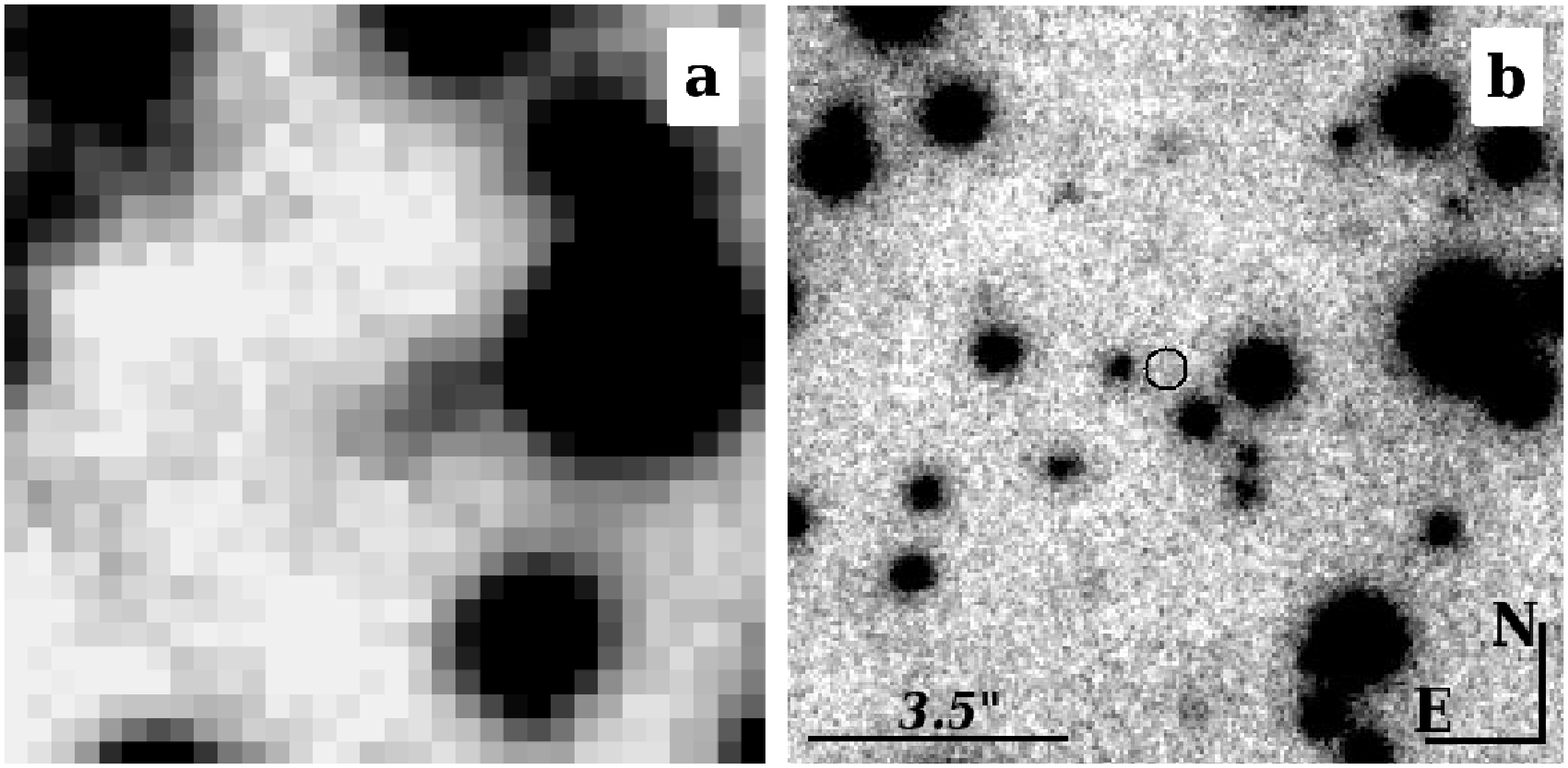}
\caption{}
\end{figure}

\begin{figure}
\includegraphics[width=15.0cm,angle=0,clip=true]{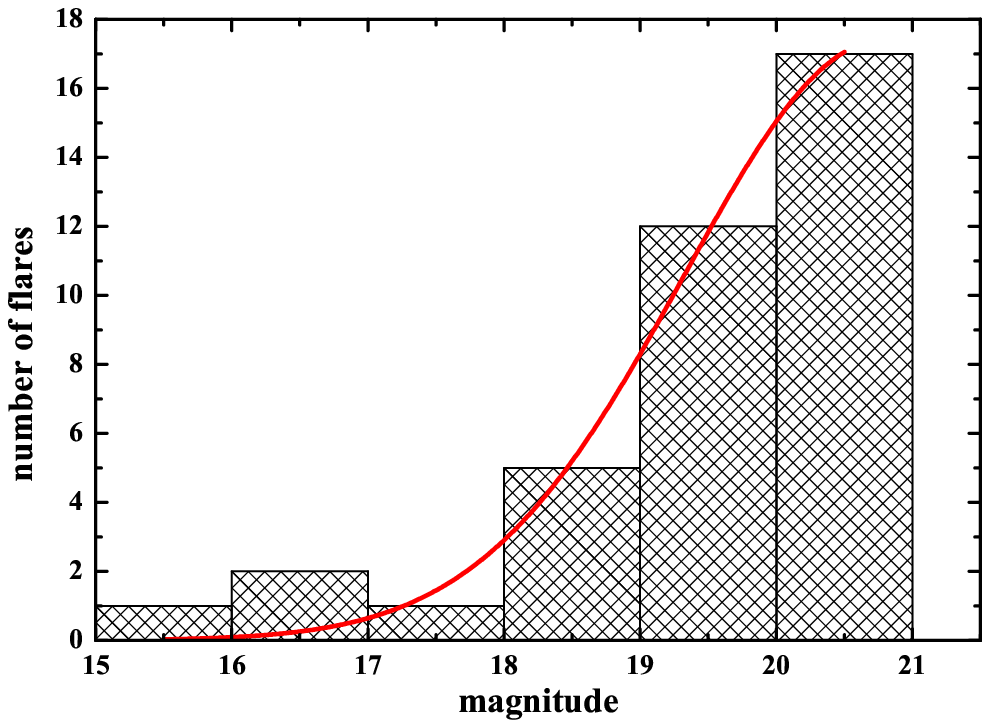}
\caption{}
\end{figure}


\end{document}